\documentclass[english,aps,prl,twocolumn,superscriptaddress,showpacs,amsmath]{revtex4-1}

\usepackage[T1]{fontenc}
\usepackage[latin9]{inputenc}
\usepackage{color}
\usepackage{pdfpages}
\usepackage{amssymb}
\usepackage{graphicx}
\usepackage{esint}
\usepackage{bm}
\usepackage{natbib}
\usepackage[normalem]{ulem}
\usepackage[breaklinks=true,colorlinks=true,urlcolor=blue,
citecolor=blue,linkcolor=blue,bookmarks=false]{hyperref}

\usepackage{babel}
\usepackage{ulem}
\makeatletter
\AtBeginDocument{\let\LS@rot\@undefined}
\makeatother

\begin{document}

\title{Large Landau level splitting with tunable one-dimensional graphene superlattice probed by magneto capacitance measurements}

\author{Manabendra Kuiri}
\affiliation{Department of Physics, Indian Institute of Science, Bangalore 560012, India}
\author{Gaurav Kumar Gupta}
\affiliation{Department of Physics, Indian Institute of Science, Bangalore 560012, India}
\author{Yuval Ronen}
\affiliation{Braun Center for Submicron Research, Department of Condensed Matter Physics, Weizmann Institute of Science, Rehovot 76100, Israel}
\affiliation{Department of Physics, Harvard University, Cambridge, MA 02138, USA}
\author{Tanmoy Das}
\affiliation{Department of Physics, Indian Institute of Science, Bangalore 560012, India}
\author{Anindya Das}
\affiliation{Department of Physics, Indian Institute of Science, Bangalore 560012, India}
\email{anindya@physics.iisc.ernet.in}

\date{\today}


\begin{abstract}
The unique zero energy Landau Level of graphene has a particle-hole symmetry in the bulk, which is lifted at the boundary leading to a splitting into two chiral edge modes. It has long been theoretically predicted that the splitting of the zero-energy Landau level inside the {\it bulk} can lead to many interesting physics, such as quantum spin Hall effect, Dirac like singular points of the chiral edge modes, and others. However, so far the obtained splitting with high-magnetic field even on a hBN substrate are not amenable to experimental detection, and functionality. Guided by theoretical calculations, here we produce a large gap zero-energy Landau level splitting ($\sim$ 150 meV) with the usage of a one-dimensional (1D) superlattice potential. We have created tunable 1D superlattice in a hBN encapsulated graphene device using an array of metal gates with a period of $\sim$ 100 nm. The Landau level spectrum is visualized by measuring magneto capacitance spectroscopy. We monitor the splitting of the zeroth Landau level as a function of superlattice potential. The observed splitting energy is an order higher in magnitude compared to the previous studies of splitting due to the symmetry breaking in pristine graphene. The origin of such large  Landau level spitting in 1D potential is explained with a degenerate perturbation theory. We find that owing to the periodic potential, the Landau level becomes dispersive, and acquires sharp peaks at the tunable band edges. Our study will pave the way to create the tunable 1D periodic structure for multi-functionalization and device application like graphene electronic circuits from appropriately engineered periodic patterns in near future.
\end{abstract}
\maketitle

\section{Introduction}
One dimensional (1D) periodic superlattice potential has been a subject of considerable experimental and theoretical interest in condensed matter physics for decades\cite{tsu1973tunneling,tsu2010superlattice}. Graphene and its superlattice structures like Moire pattern\cite{PhysRevLett.102.056808,yankowitz2012emergence} have emerged a new play ground in recent years for both fundamentally due to its chiral nature as well as application point of view. Interesting electronic and transport phenomena such as, Klein tunnelling\cite{katsnelson2006chiral}, electron collimation\cite{park2008electron}, lensing\cite{cheianov2007focusing}, emergence of cloned Dirac points\cite{ponomarenko2013cloning,hunt2013massive}, self similar recursive energy spectrum such as Hofstadter butterfly\cite{dean2013hofstadter,hunt2013massive}, topological phases like floquet topological insulator \cite{PhysRevB.90.115423} and different topological aspects, such as change in winding number as well as different scattering mechanisms \citep{wang2015topological} have been demonstrated based on graphene superlattice.\\

There have been several theoretical studies on 1D graphene superlattice and one of the main findings is that 1D periodic potential in graphene will give rise to new type of Dirac cones at zero energy as well as anisotropic Fermi velocity renormalization\citep{park2008anisotropic,PhysRevLett.103.046809,PhysRevB.81.205444,tan2011new,PhysRevB.81.075438,PhysRevLett.107.086801}. Some of them have been observed experimentally by looking at the appearance of the multiple resistance peaks as a function of superlattice potential\cite{dubey2013tunable,PhysRevB.89.115421,drienovsky2017few,forsythe2017band}. However, contradictory conclusions have been drawn about the origin of multiple resistance peaks. No measurements have been performed to probe directly the renormalization of 1D superlattice band. Thus, direct probe of the superlattice band together with complete theoretical understanding of 1D periodic potential in graphene is crucial as it can shed light on the electronic properties as well as desirable for device applications.\\

The recent theoretical studies of 1D superlattice potential on the Landau level (LL) spectrum of graphene, in particular its effect on the zeroth LL has drawn considerable interest in the community\cite{PhysRevLett.103.046808,PhysRevB.85.195404,PhysRevB.85.235457,jappaper}. The unique zeroth LL in pristine graphene having particle-hole symmetry gives rise to anomalous quantum Hall sequence compared to conventional semiconductor. It has been demonstrated both theoretically\cite{PhysRevLett.96.176803} as well as experimentally\cite{PhysRevLett.96.136806,PhysRevLett.99.106802,PhysRevLett.100.206801,bolotin2009observation} that the degeneracy of the N = 0 LL of pristine graphene can be lifted by breaking the symmetries at high magnetic field, which gives rise to the splitting of the zeroth LL\cite{PhysRevLett.96.136806,PhysRevLett.99.106802,PhysRevLett.100.206801,bolotin2009observation} and manifests many interesting physics like quantum spin Hall state or helical state. Now, it has been shown theoretically that the application of 1D superlattice potential will change the degeneracy of the zeroth LL ($N=0$) as well as the Hall conductivity step sequence with the creation of new Dirac cones \cite{PhysRevLett.103.046808}. It has been pointed out in ref \cite{PhysRevB.85.195404} that the effect of 1D superlattice on LL spectrum will depend on the competition between the magnetic length scale ($l_{B}$) versus period of the 1D potential ($L$). In the weak limit ($l_{B}$ > $L$) it will only change the degeneracy of the levels. However, in the intermediate ($l_{B}$ $\sim$ $L$) and strong limit ($l_{B}$ < $L$) the LL becomes dispersive and the ref \cite{PhysRevB.85.235457} predicted a large energy splitting of the $N=0$ Landau level (LL) comparable to the strength of the 1D supperlattice potential. However, there are no experimental observation of the renormalization of the LL spectrum in presence of 1D superlattice potential till date.\\

\begin{figure*}[ht!]
 \includegraphics[width=0.8\textwidth]{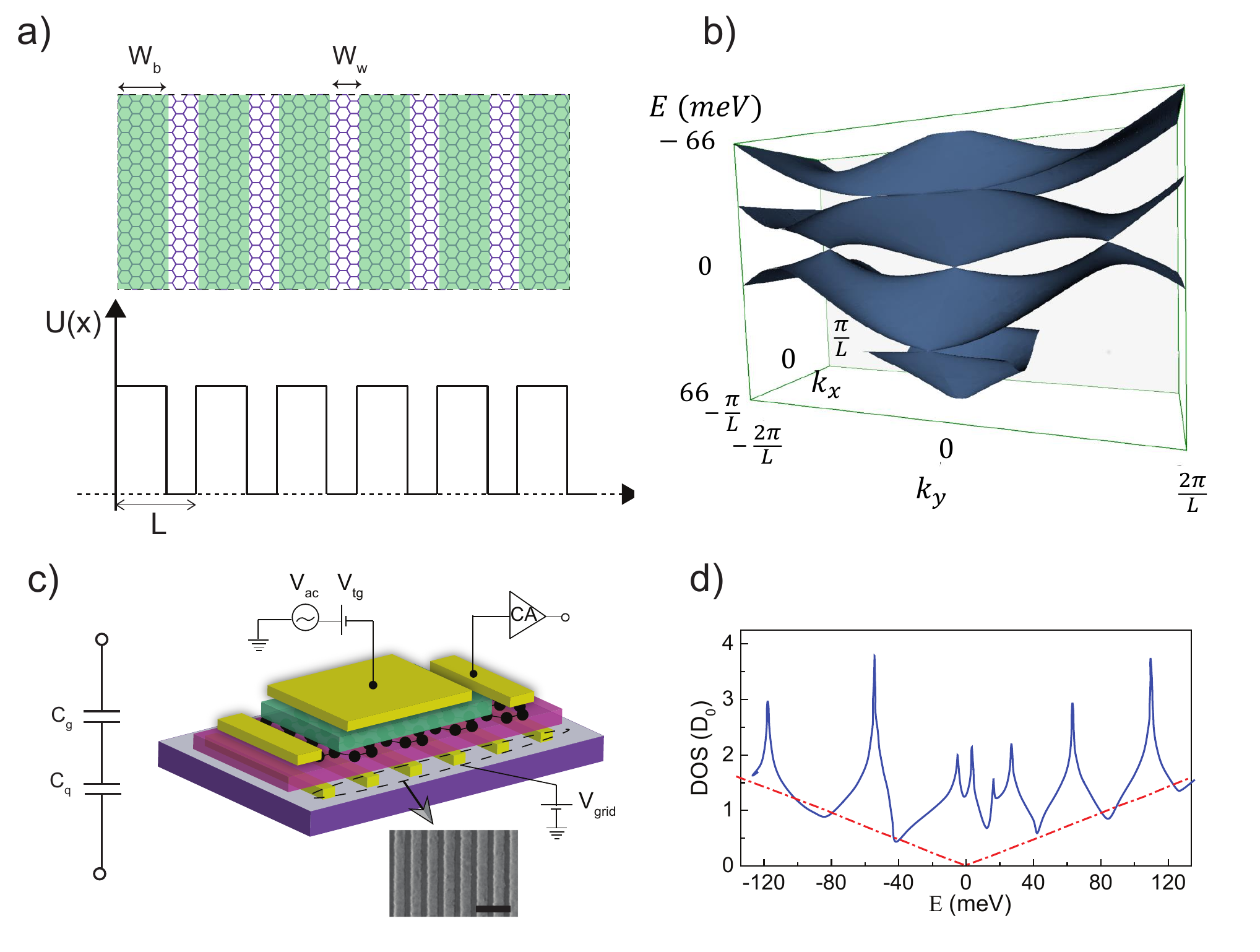}
 \caption{(Color Online) (a) Schematic of one dimensional periodic potential. The periodic potential is applied along the $x$ direction with a potential strength of $U_0$. $W_b$ and $W_w$ correspond to the regions of potential barrier and no barrier, respectively. (b) Band dispersion of a graphene superlattice with barrier width $W_b=0.6$ and $W_w=0.4$ and $U=130$ meV. (c) Schematic of the device architecture. Bottom inset shows the scanning electron microscope image of the grids. The scale bar is $200~nm$. Left inset shows equivalent capacitance model between the top gate and graphene. (d) DOS for $W_b=0.5$ and $W_w=0.4$ adopted from Ref \cite{PhysRevB.81.075438} (blue). The DOS without superlattice potential is shown in dashed red line.}
 \label{fig:example}
\end{figure*}

Quantum capacitance\cite{luryi1988quantum} measurements reveals significant insights to {\it e-e} interactions\cite{PhysRevLett.68.674,ilani2006measurement,yu2013interaction,chen2013electron,zibrov2017tunable}, quantum correlation\cite{ilani2006measurement}, layer polarization\cite{PhysRevB.85.235458,hunt2017direct}, many body physics\cite{ilani2006measurement,zibrov2017tunable,hunt2017direct}, Fermi velocity renormalization\cite{yu2013interaction,chen2013electron}
 and thermodynamic compressibility\cite{PhysRevLett.105.136801,droscher2010quantum,PhysRevB.85.235458,chen2013electron} in graphene. As the quantum capacitance is directly proportional to thermodynamic density of states (DOS), making it possible to directly probe the renormalization of 1D superlattice band\cite{PhysRevB.92.075408,wang2013negative}. In this report, we have carried out resistance and quantum capacitance measurements on a boron nitride (hBN) encapsulated graphene device while 1D periodic superlattice potential and magnetic field applied. Our main observations are the following: (i) Inducing 1D periodic potential give rise to multiple resistance peaks at the Dirac point, which are reflected in the quantum capacitance data, consistent with the theoretical model. (ii) Magneto capacitance measurements have revealed two prominent features: at low superlattice potential $U$ $<$ 100 meV, broadening of $N=0$ LL as well as other LL was observed. At higher superlattice potential $U$ $>$ 100 meV, we observed evolution of the $N=0$ LL. First splitting into two sub levels, further increase in superlattice potential it starts interacting with the nearest Landau level. The observation of one order larger energy splitting ($\sim 150$ meV) of the $N=0$ LL with the application of 1D superlattice potential is completely new emerging phenomena as compared to the observed splitting of zeroth LL at high magnetic field ($10-15$ meV) due to symmetry breaking in pristine graphene\cite{PhysRevLett.96.136806}. Our observations are compared with our corresponding theoretical calculations based on degenerate perturbation theory. We find that the Landau level becomes dispersive with the application of 1D superlattice potential and acquires sharp peaks at the tunable band edges exhibiting zeroth LL splitting.\\

\section{1D superlattice and band structure}
Fig.~1a shows the schematic of graphene placed on 1D artificial periodic superlattice. Here the periodic potential is applied along the $x$ direction with potential strength $U$. Due to the chiral nature of charge carriers in graphene the group velocity does not change along the direction of applied potential, rather the group velocity gets renormalized along the direction perpendicular to the applied potential. It has also been shown that at the reduced zone boundary (MZ) there will be a band gap opening at the finite value of $k_y$ and its value will depend on the strength of the periodic potential \cite{park2008anisotropic}. The solution for square barrier superlattice is given as \cite{PhysRevB.81.075438}
\begin{equation}\label{1}
	\cos k_x = \cos\lambda W_w \cos\Lambda W_b - G \sin\lambda W_w \sin\Lambda W_b.
\end{equation}
where $G=(\epsilon_w \epsilon_b -k_y^2)/\lambda\Lambda$, $\lambda=(\epsilon_w^2-k_y^2)^{1/2}$, $\Lambda=(\epsilon_b^2-k_y^2)^{1/2}$, $\epsilon_w=\epsilon+UW_b$, $\epsilon_b=\epsilon-UW_w$; $W_b \text{ and }W_w$ are the width of the barrier and the well, respectively, $\epsilon$ is the energy and $U$ is the strength of the grid potential. The band structure calculated using this equation is shown in Fig.~1b with $W_b=0.6$ and $W_w=0.4$ for $U$ $\sim$ 130meV and the generated DOS is shown in Fig. 1d (\cite{PhysRevB.81.075438}), where the dashed red line shows the DOS for zero superlattice potential. It can be seen from the Fig. 1d that dip in the DOS arises at the zone boundary; G (reciprocal lattice vector) = 2$\pi$/L $\sim$ 40 meV, where L is the super lattice length ($\sim$ 100 nm).

\begin{figure*}[ht!]
 \includegraphics[width=0.8\textwidth]{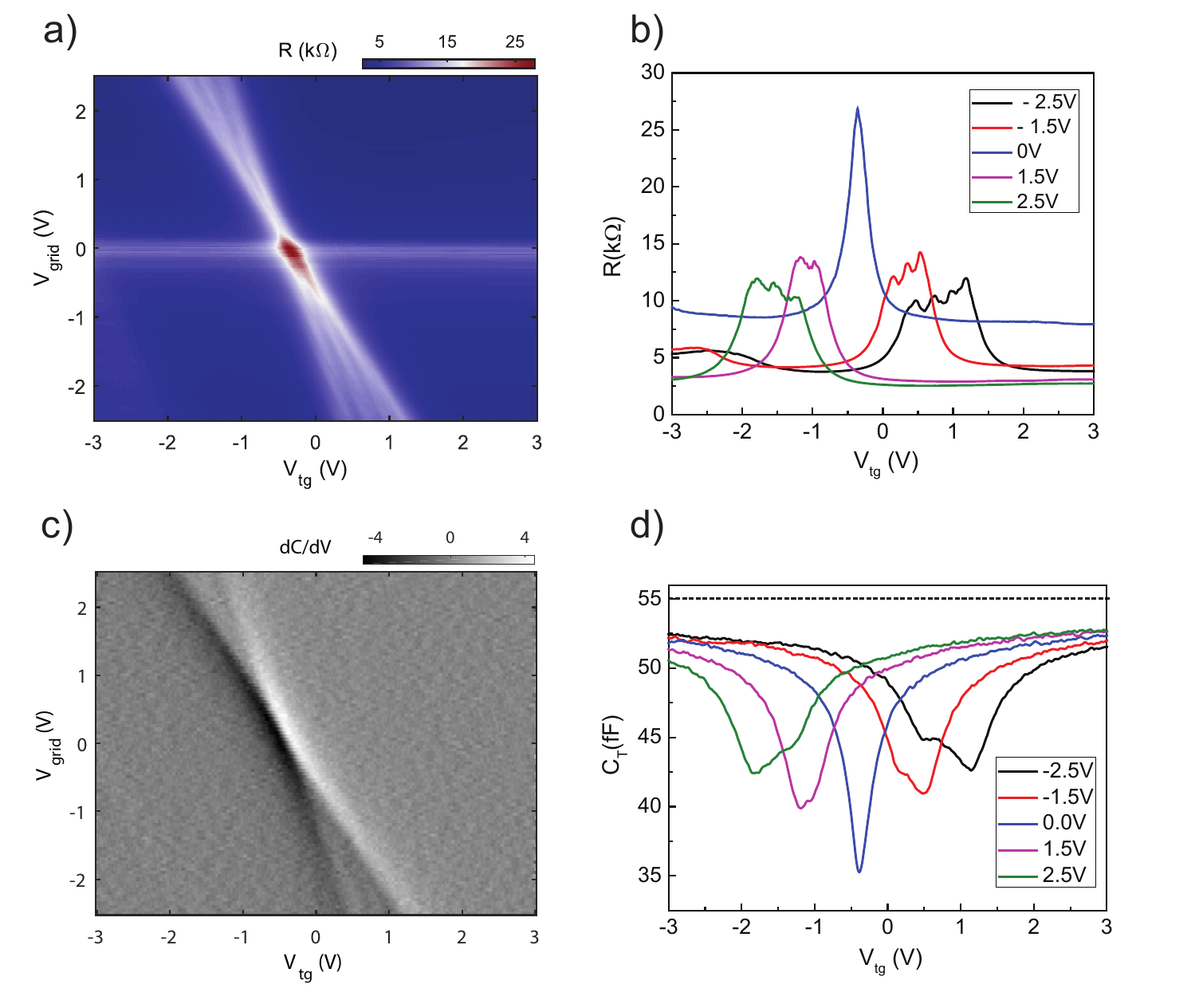}
 \caption{(Color Online) (a) Colorplot of two probe resistance as a function of topgate voltage ($V_{tg}$) and grid voltage ($V_{grid}$) at $T=2K$. (b) Resistance cut lines for $V_{grid}=0V,~\pm1.5V,~\pm2.5V$. (c) Derivative of the total measured capacitance as a function of topgate voltage ($V_{tg}$) and grid voltage ($V_{grid}$). (d) Total capacitance as a function of topgate voltage($V_{tg}$) for grid voltage $V_{grid}=0V,~\pm1.5V,~\pm2.5V$. The dashed horizontal line is the geometrical capacitance between the top gate and graphene.}
 \label{fig:example}
\end{figure*}

\section{Methods and measurement techniques}
The schematic of the device architecture is shown in Fig.~1c. To realize 1D superlattice structures in graphene, we fabricated the device in the following way. At first Si/SiO$_2$ substrate  with 150 nm oxide was  used to design the periodically spaced gold gates with barrier width $W_b=60$ nm and well $W_w=40$ nm using electron beam lithography followed by thermal deposition of Ti/Au (5/20) nm. The periodicity in this case was $L \sim$100 nm. The inset in Fig.~1c shows the Scanning electron microscope image (SEM) of the gold grids. The details have been given in the supplementary information (S.I). Once the grids were ready, the top hexagonal boron nitride(hBN) and graphene were mechanically exfoliated separately onto piranha cleaned Si/SiO$_2$ substrate and picked up individually one after another following the similar method as mentioned in Ref\cite{wang2013one,zomer2014fast}. The bottom hBN was also picked up using the same procedure and the final hBN-graphene-hBN stack was transferred onto the gold grids as shown schematically in Fig.~1c (S1 for details). Top gate and source(drain) electrodes were patterned using electron beam lithography followed by thermal deposition of Cr(5nm)/Au(70nm). All the measurements are carried out Oxford $He^3$ cryostat at T = 2K. The capacitance was measured between the top gate and graphene.
 A small ac voltage of $\sim (10- 15)$ mV at a frequency of $\sim$ 5 kHz was applied to the top gate and the out of phase current was measured by lock in amplifier using home build current amplifier with a gain of 10$^7$\cite{kretinin2012wide} as described in our previous work \cite{kuiri2015probing}. The area of the top gate was $\sim$ 18 $\mu m^2$ (determined by optical microscopy as well as by scanning electron microscopy) and the thickness of the top BN was $d_{t}\sim$ 10-11 nm (determined by atomic force microscopy) yielding an effective top gate capacitance, C$_g$ of $\sim$ 55 $fF$ (for $d_{t}$ $\sim$ 10.5 nm) (S2-S4 for details). 

\section{Resistance Data}
Two probe resistance measurement were performed using low frequency lock in technique by passing a small voltage between source-drain ($\sim$ 100 $\mu V$) and measuring the current through the channel. Fig.~2a  shows the two terminal resistance color plot of the device as a function of grid voltage (V$_{grid}$) and top gate voltage (V$_{tg}$), where the carrier density in graphene was tuned globally using periodic gold gate V$_{grid}$ and locally using V$_{tg}$. It can be seen that the Dirac point is located at (V$_{grid}$,V$_{tg}$) = (0.0V, - 0.5V). For V$_{grid}$ = 0V, we can see the usual resistance gate voltage characteristic with V$_{tg}$ as shown as a cut line in Fig.~2b. With the increase in superlattice potential i.e. at finite V$_{grid}$, we see the appearance of additional peaks; for V$_{grid}$ = 1V we see two peaks, for V$_{grid}$ = 2.5V we see three peaks in the resistance data. The corresponding cut lines  of Fig. 2a are shown in Fig. 2b for V$_{grid}$ = $\pm$1.5V and $\pm$2.5V and it can be seen that the additional resistance peaks appeared symmetrically for both positive and negative superlattice potentials.\\ 

The application of the V$_{grid}$ makes an unequal Fermi energy (E$_{F}$) shifts in graphene; the E$_{F}$ shift will be more on top of the gold pillar ($W_b=60$ nm) compared to the gap part ($W_w=40$ nm). Thus, there will be a barrier between the two parts. 
The potential barrier, $U$, defined by both gates $V_{tg}$ and $V_{grid}$ is given by (assuming equal barrier and well width $W_b=W_w$)\cite{dubey2013tunable}
\begin{equation}
U= \sqrt{\pi}\hbar v_F\left(\sqrt{\frac{C_{grid}\Delta V_{grid}+{C_{tg}\Delta V_{tg}}}{e}} - \sqrt{\frac{C_{tg}\Delta V_{tg}}{e}} \right)
\end{equation}
The only unknown parameter of the above equation is $C_{grid}$, which was estimated from the diagonal line of the charge neutrality point in Fig. 2a as described in SI (S5 for details).
The calculated barrier heights for V$_{grid}$ = 1.0V, 1.5V, 2.0V and 2.5V are 100 meV, 122 meV, 140 meV and 165 meV, respectively. We would like to comment that the above mentioned values are slightly overestimated as we have not considered the asymmetric grid potential as well as fringing effect of the electric field\cite{dubey2013tunable}.\\

It is evident that due to the super-lattice potential the multiple resistance peaks appear in Fig. 2a and 2b, which could be due to the dip in the DOS 
as shown theoretically in Fig. 1d 
and observed experimentally in Ref \cite{dubey2013tunable}. However, the Ref \cite{PhysRevB.89.115421} claims that the origin of multiple resistance peaks is due to Fabry Perot resonance. In order to see whether the multiple resistance peaks are indeed an effect of DOS modification or not we will now present our quantum capacitance data.

\begin{figure}[t!]
 \includegraphics[width=0.4\textwidth]{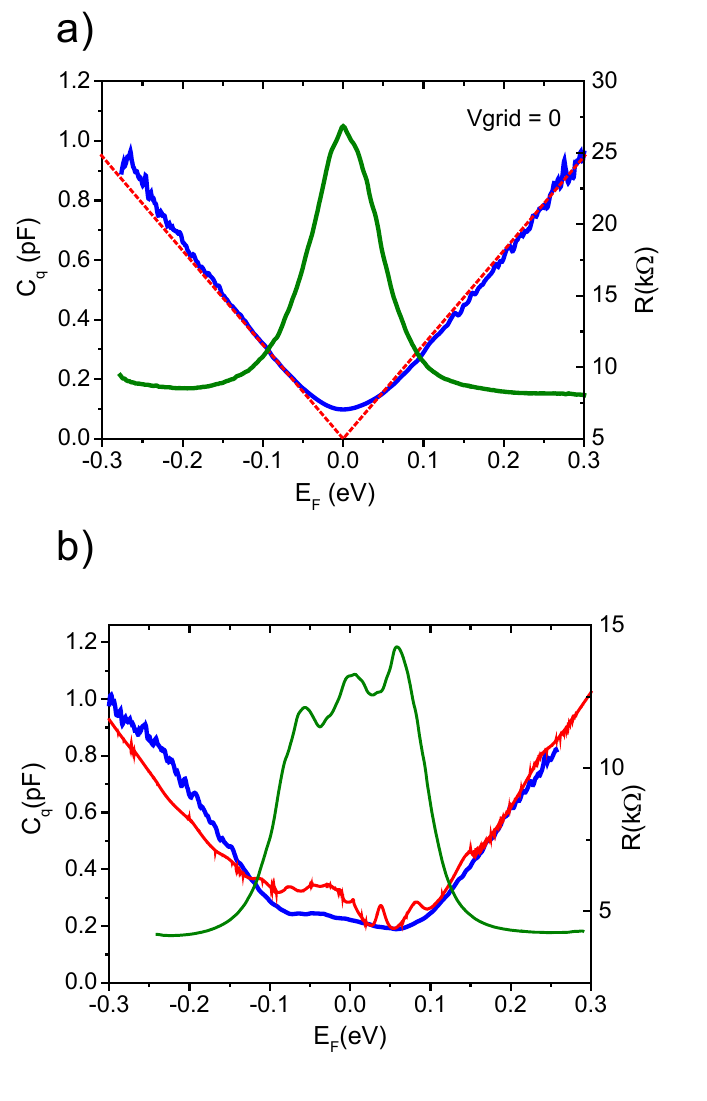}
 \caption{(Color Online) (a) Blue solid curve shows the extracted quantum capacitance as a function of Fermi energy $E_F$ for $V_{grid}=0 V$. The red dashed line shows the single particle quantum capacitance versus $E_F$ of monolayer graphene with the experimental value of $C_g=55 fF$. The olive solid line shows the resistance plot with $E_F$ (b) Blue solid line shows the quantum capacitance as a function of $E_F$ for $V_{grid}= -2V$. Red line shows the theoretical DOS as a function $E_F$, calculated for asymmetric potential with $W_b=0.6$ and $W_w=0.4$ as described by Ref \cite{PhysRevB.81.075438}, with added Fermi energy broadening of $\sim 40~meV$.}
 \label{fig:example}
\end{figure}

\section{Capacitance Data for B=0}
In a parallel plate capacitor made of graphene and a normal metal, adding an electron to the graphene layer will cost not only the electrostatic potential energy but also kinetic energy. As a result, the measured capacitance will have both the geometrical term (potential) as well as the quantum term (kinetic) \cite{luryi1988quantum}. The total measured capacitance in such a system is given by 

\begin{equation}
C_t= \left(\frac{1}{C_g} + \frac{1}{Ae^2\frac{dn}{d\mu}} \right)^{-1} + C_p   
\label{eqn:eq1}
\end{equation}

\begin{figure}[t!]
\includegraphics[width=0.4\textwidth]{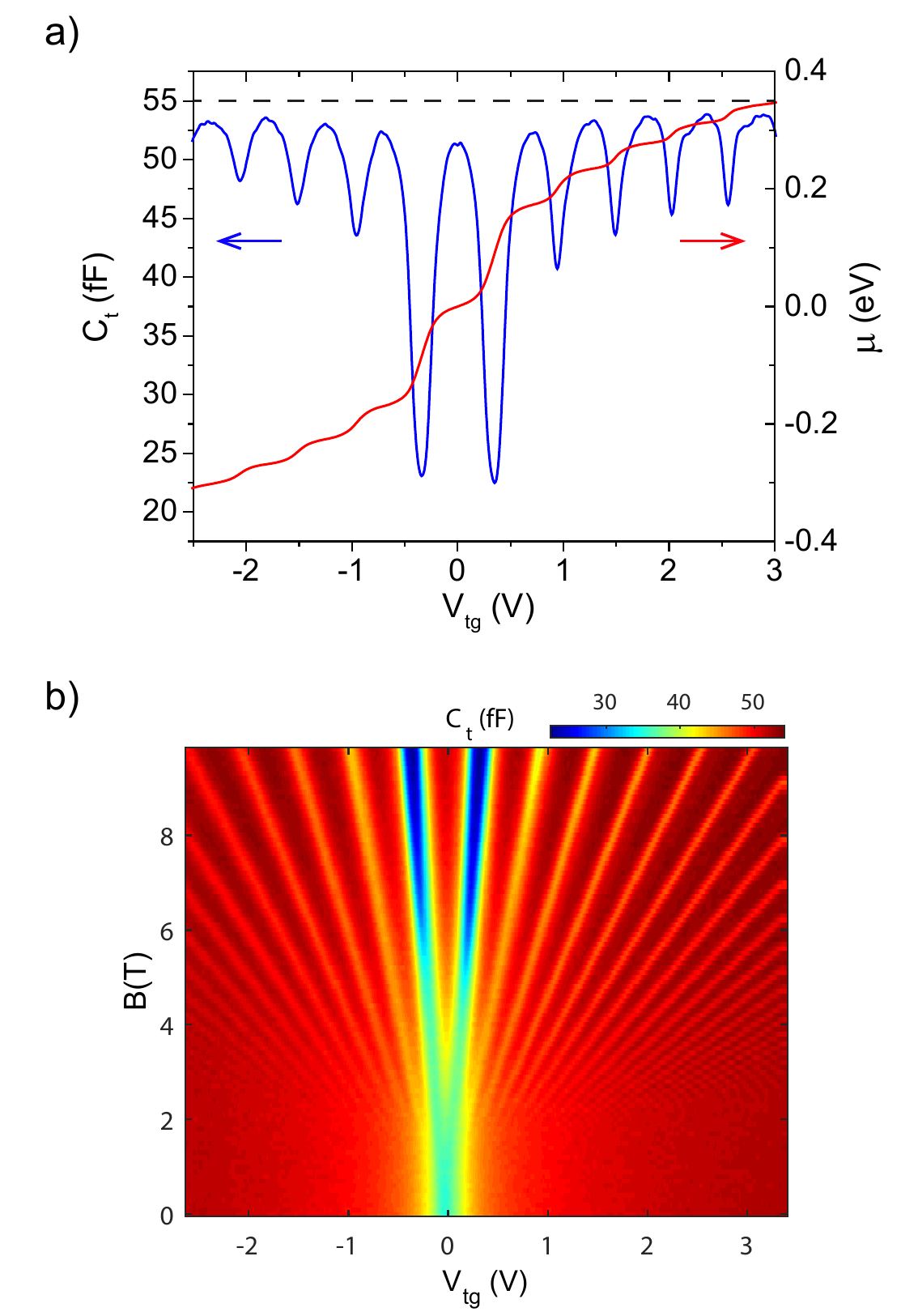}
\caption{(Color online) (a) Measured total capacitance($C_t$) as a function of topgate voltage($V_{tg}$) at $B=9.8T$ (blue line). The red line shows the corresponding change in chemical potential as a function of $V_{tg}$ using the charge conservation relation.
(c) Fan diagram for $V_{grid}=0V$, where we plot $C_t$ as a function of $V_{tg}$ for different $B$}
\label{Fig1:band}
 \end{figure}
 
where, $C_g$ is the geometrical capacitance between the top gate and graphene, $A$ is the area of the graphene capacitor, and $\frac{dn}{d\mu}$ is the thermodynamic compressibility \cite{yu2013interaction}. C$_p$ is the parasitic capacitance associated with wiring plus stray capacitance. The quantum capacitance is defined as $C_q = Ae^2\frac{dn}{d\mu}$, which is directly proportional to DOS. The differential capacitance was measured using the similar techniques in Ref \cite{kuiri2015probing}. The parasitic capacitance (C$_p$ $\sim$ 4.8 $fF$) arising due to the electrical wiring and stray capacitance has been subtracted from the measured capacitance data. The details about the determination of parasitic capacitance ($C_p$) is shown in S3. Fig. 2c shows the 2D color plot of derivative of the total capacitance ($C_t$) measured as a function of $V_{tg}$ and $V_{grid}$, where one can see the multiple broadened dips with increasing superlattice potential. Although the derivative of the capacitance shows three weaker dips at higher superlattice potential but the data is not as prominent as the multiple resistance peaks seen in Fig. 2a. Fig. 2d shows the cut lines of total capacitance ($C_t$) as a function of $V_{tg}$ for several grid potentials.\\

In order to understand the experimental capacitance data, we need to plot the quantum capacitance (C$_q$) as a function of the Fermi energy ($E_F$). The $C_q$ can be calculated as $C_q^{-1}=C_t^{-1}-C_g^{-1}$ and $E_F$ can be calculated from the charge conservation relation \cite{xu2011quantum,droscher2010quantum,yu2013interaction} $E_F=e\int_{0}^{V_{tg}}\left(1-\frac{C_t}{C_g}\right)dV_{tg}$, where $e$ is the electronic charge. Fig.~3a shows the $C_q$ vs $E_F$ plot for zero grid voltage. The linear nature of the curve is reflecting the linear DOS of graphene. Using the DOS of graphene, $D(\mu)$ = $\frac{2|\mu|}{\pi (\hbar v_F)^2}$ and $v_F=1.2\times 10^6 m/s$ is the Fermi velocity, we generate the theoretical curve in Fig. 3a (red dotted line). It can be noticed that the experimental $C_q$ does not vanish at $E_F=0$ signifying the residual charge in-homogeneity in the device and we estimate the Fermi energy broadening ($\Delta E_F\sim 30-40\;meV$). $C_q$ vs $E_F$ plot for $V_{grid}= -2V$ ($U \sim 122$ meV) is shown in Fig.~3b. It can be seen how the DOS is modified near the Dirac point. The red line is the theoretically calculated DOS, 
with $\Delta E_F\sim 40\;meV$, 
where sharp features of theoretical DOS in Fig. 1d are broadened and one can notice that for $E_{F} > U = 122~meV$ (V$_{grid} = -2V$) the DOS captures the expected linear nature. It can be seen the qualitative agreement between the experimental and theoretical data in Fig. 3b. Although the positions of the extreme two resistance peaks match exactly to the positions of the extreme two dips in C$_q$ data but the signature of the middle peak in the resistance data is not evident in C$_q$. This discrepancy could be understood as follows: the resistance measurement is a directional dependent phenomena compared to the total DOS (C$_q$) measurement.\\ 

\begin{figure*}[ht!]
 \includegraphics[width=0.7\textwidth]{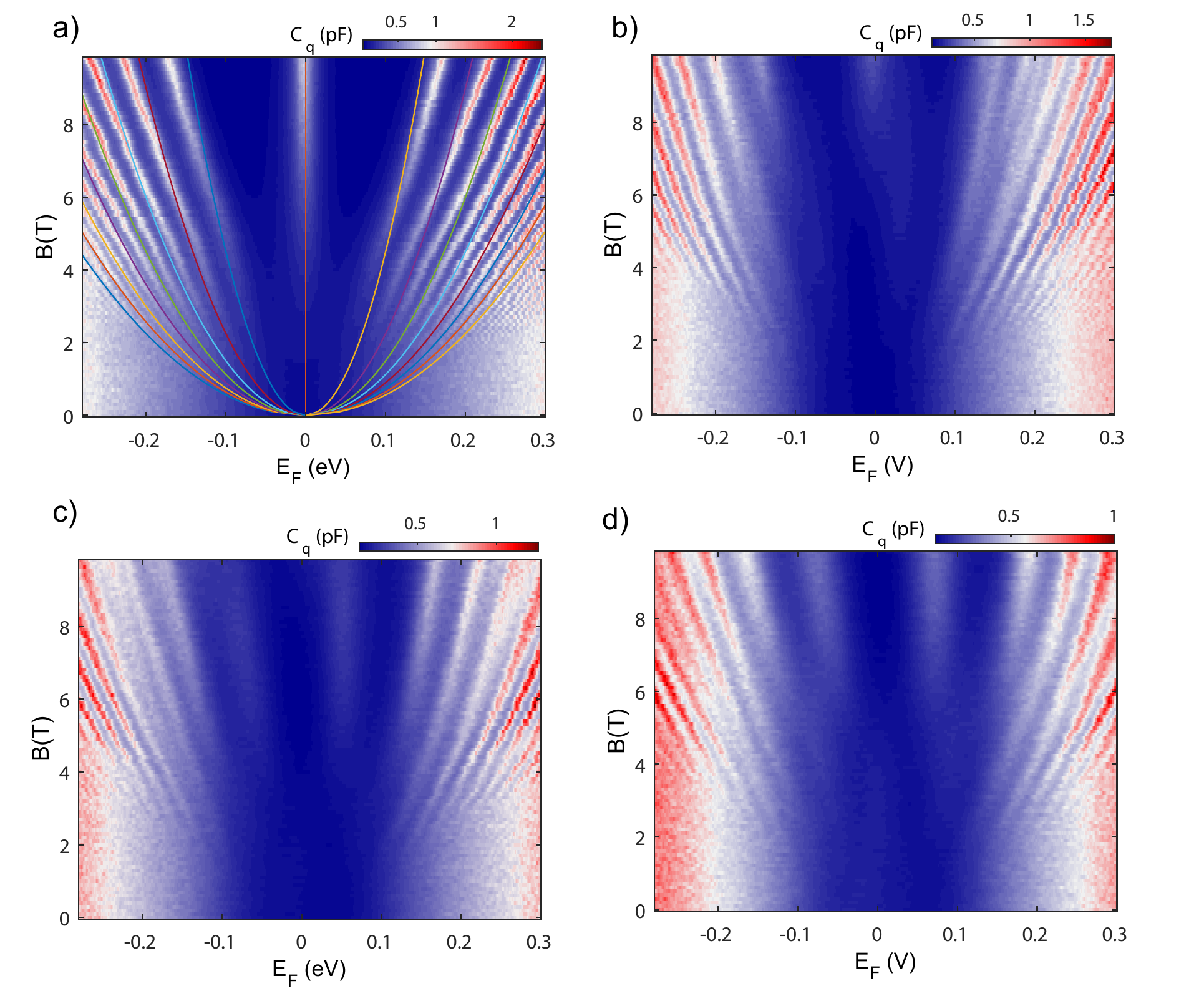}
 \caption{(Color Online) (a) The Landau level spectrum (measured quantum capacitance) for $V_{grid}=0V$ as a function of magnetic field. The solid line corresponds to the theoretically predicted one. (b) $C_q$ vs $E_F$ as function of magnetic field for $V_{grid}=-1.0V$. (c) $C_q$ vs $E_F$ as function of magnetic field for $V_{grid}=-1.5V$.(d) $C_q$ vs $E_F$ as function of magnetic field for $V_{grid}=-2.0V$.}
 \label{fig:example}
\end{figure*}
As shown in Ref \cite{park2008anisotropic} that the application of super-lattice potential along the $x$ direction cannot open a gap along that direction ($k_x$) due to the Klien tunneling. Also there will be no gap in the perpendicular direction ($k_y$). However, there will be gap opening at the reduced zone boundary in the certain directions of $k_x-k_y$ plane. It can be seen from our SEM image of the device (S4), the graphene is placed at an angle of $\sim$ 60 $^{\circ}$ with the super-lattice direction. Thus, we believe that our measured longitudinal resistance predominantly capture the DOS along a certain direction of $k_x-k_y$ plane where as the quantum capacitance measurement captures the total DOS. We should note that the average separation of the resistance peaks in Fig. 3b is $\sim$ 50 meV. This energy separation is close to the expected DOS modifications at the reduced zone boundary position; E$_Z$ = $\hbar v_F$G = $\hbar v_F$2$\pi$/L $\sim$ 40meV for super-lattice length, L = 100nm.

\section{Magneto-capacitance data}

In order to see the effect of super-lattice potential on Landau Level (LL) spectrum we have carried out the quantum capacitance measurement in perpendicular magnetic field. In the presence of perpendicular magnetic field, the energy levels in monolayer graphene is quantized, with energy eigenvalues $E_N=\pm\sqrt{2e\hbar v_F^2|N|B}$, where $|N|$ = 0, $\pm$1, $\pm$2,...is the Landau level index. Fig. 4a shows the total capacitance ($C_{t}$) as a function of $V_{tg}$ at B = 9.8T for $V_{grid}$ = 0V. The dips in the $C_t$ corresponds to the Landau-level gap. The Fig. 4b presents the total capacitance in a 2D color plot for several B showing the evolution of several Landau-levels. Although we have measured the $C_{g}$ by measuring the thickness of top hBN ($d_{t}$) by AFM and area ($A$) by SEM, but one can cross check that value from the Landau-level spectrum (Fig. 4b). First of all, the spacing (in gate voltage) between adjacent capacitance minimum around N = 0 LL remains similar to the spacing around the $N = \pm$1 LL confirming the sample is monolayer graphene as the LL degeneracy remains same for $|N|$ = 0, $\pm$1, $\pm$2, $\pm$3. The no of electron to fill the N = 0 LL is $n = 4BA/(h/e)$, where 4 comes from the 2 valley and 2 spin degrees of freedom. In the first order approximation we can write $n = C_{g} \Delta V_{tg}/e = 4BA/(h/e)$, where the $\Delta V_{tg}\sim 0.53$ is the average spacing between the adjacent $C_{t}$ minimum. 
The above relation gives $C_{g}$ $\sim$ 52$fF$, which is similar to the value obtained from measuring the thickness of top hBN (55 $fF$) by AFM. As mentioned before using the charge conservation equation ($E_F=e\int_{0}^{V_{tg}}\left(1-\frac{C_t}{C_g}\right)dV_{tg}$), we have converted $V_{tg}$ into $E_{F}$ or $\mu$, which is shown in Fig. 4a, where one can clearly see the non-monotonic increment of $\mu$ with $V_{tg}$ due to the presence of compressible and in-compressible regions in LL spectrum. Using the conversion method we have plotted the LL energy spectrum of graphene at zero super-lattice potential in Fig. 5a, where one can see the $|N|$ = 0, $\pm$1, $\pm$2....$\pm$8 very clearly. The solid lines are the theoretical one based on $E_N=\pm\sqrt{2e\hbar v_F^2|N|B}$. Agreement between the experimental data with the theory is quite noticeable upto $\sim$ 6T. Fig. 5b, 5c and 5d presents the LL spectrum for superlattice potentials, $U = 100 ~meV, 122 ~meV$ and $140 ~meV$, respectively. At $U = 100 ~meV$ the zeroth LL gets broadened. However, the most noticeable changed was observed beyond $U = 122 ~meV$, where the N = 0 LL gets splitted and at higher super-lattice potential ($U>140 ~meV$) the splitted N = 0 LL starts interacting with higher LLs as shown in S8.\\

The zeroth LL in monolayer graphene gives rise to anomalous quantum Hall sequence\cite{PhysRevLett.96.176803} in graphene compared to conventional semiconductor. However, it is known from the literature\cite{PhysRevLett.96.136806,PhysRevLett.99.106802,PhysRevLett.100.206801,bolotin2009observation} that the N = 0 LL degeneracy can be lifted by breaking the symmetries at high magnetic field, which gives rise to an insulating state\cite{checkelsky2009divergent,PhysRevLett.99.106802,bolotin2009observation} at the Dirac point. The common consensus is that the valley degeneracy gets lifted at higher B due to the coulomb interaction ($e^2/l_B$, where $l_B=(\hbar/eB)^{1/2}$ is the magnetic length scale), which is a many body interaction physics. However, in order to see the insulating phase at $N = 0$ LL ($\Delta E_{I} \sim 10-15$ meV at 10T) one needs to have extremely cleaner device ($\Delta E_{I}>\Delta E_{F}$). As can be seen from Fig. 5a that we do not observe the splitting of the zeroth LL even upto 10T at U = 0 because of large $\Delta E_{F} \sim 40$ meV in our device, but it gets splitted with the application of superlattice potential and splitting is very large $\sim 150$ meV. Thus, the large splitting of $N = 0$ LL due to the 1D super-lattice potential in Fig. 5c and 5d is completely new, which has not been experimentally reported prior to this work. In the next paragraph we will discuss our theoretical calculation based on degenerate perturbation theory to calculate the DOS of the LL in the presence of 1D super-lattice potential. It turns out that indeed the zeorth LL gets broadened with a energy dispersion having a width comparable to the strength of superlatice potential and the DOS acquires an enhanced peak at the band edge resulting in a splitting around the zeroth LL.


\section{Theory}
What is so special about a 1D superlattice potential which enables large splitting of the zeroth LL, while the typical 2D superlattice potentials often fail to do so?  We first give a heuristic argument, and supplement it with a theoretical calculation. The LLs are the quantized energy $E_n$ ($n$ is the LL index) of the cyclotron orbits in real space. The cyclotron orbits are localized with Gaussian wavefunction, with its width determined by the magnetic length $l_{\rm B}$\footnote{In the Landau gauge, the cyclotron orbitals are extended in one direction, and localized in the orthogonal direction. Here, without loosing generality, we assume the 1D potential is along the direction of the LL localization, while a similar physics can be reproduced with the other assumption}. They are thus analogous to the `maximally localized Wannier orbitals' with a localization length of $l_{\rm B}$. A weak periodic potential helps mobilizing these localized cyclotron orbits, with a range of wavevectors featuring the underlying periodicity of the superlattice potential. As a result the LLs acquire band dispersion $E_n(k)$. A 1D dispersion manifests sharp peaks in the density of states (DOS) at the top and bottom of the bands, and a dip in the low-energy spectral function. On the contrary, a 2D dispersion acquires a peak in the DOS in the low-energy region due to the van-Hove singularity. In what follows, the key message is that a 1D dispersion of the LL gives the impression of a well resolved splitting of the LL in the DOS, while a 2D potential produces an opposite effect. 
 \begin{center}
\begin{figure*}[ht]
 \includegraphics[width=0.9\textwidth]{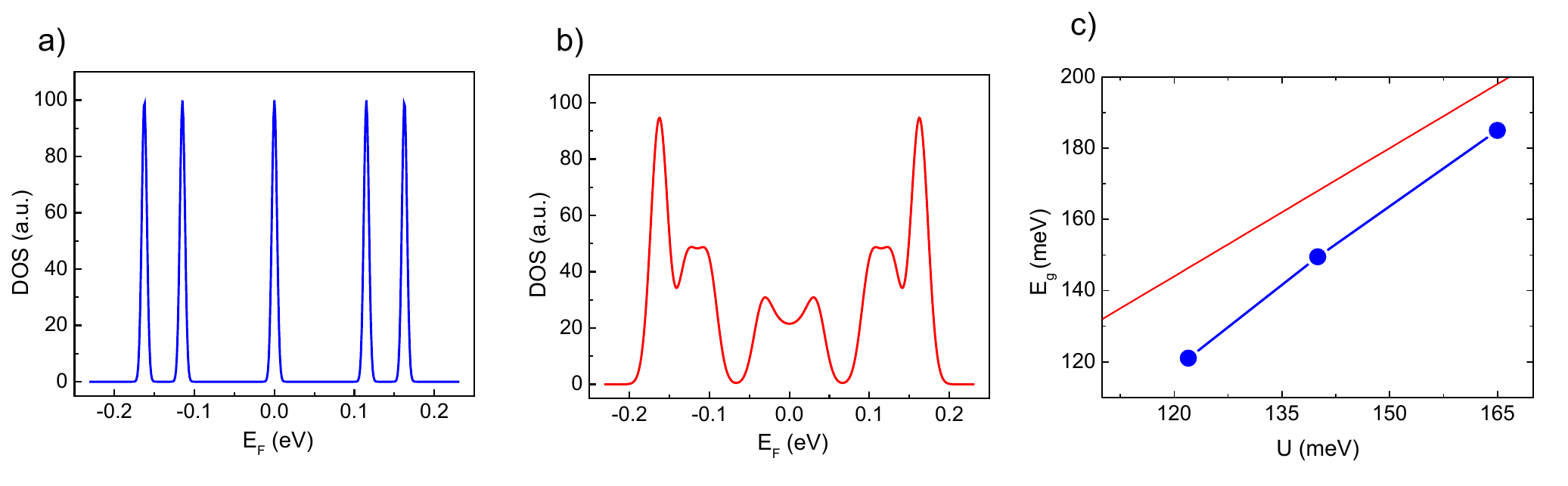}
 \caption{(Color Online) (a) Calculated DOS as a function of $E_F$ for $U=0$ meV. (b) Calculated DOS as a function of $E_F$ for $U=50$ meV. (c) Energy separation of the $N=0$ LL as a function of superlattice potential (blue). The red curve shows the theoretically calculated energy separation.}
 \label{fig:example}
\end{figure*}
\end{center}
We are in the sufficiently large magnetic field limit such that the LL spacing $\hbar\omega_c > U$, where $U$ is the superlattice potential strength. In our samples, $U$ is tunable and thus at smaller values of $U$ the inter-LL hybridization or interaction can be neglected. Owing to the large degeneracy associated with the LLs, we can therefore incorporate the effects of the weak-periodic potential within a degenerate perturbation theory for a given LL. We solve the LL problem with the Landau gauge by assuming the vector potential $A_x=-By$, which produces a wavefunction that is extended in the $y$-direction, but Gaussian like along the $x-$direction. So, we get $\psi_{n,k}({\bf r})=N_n H_{n}(\frac{y-y_k}{l_B})e^{ikx}e^{-(y-y_{k})^2/2l^2_{\rm B}}$, where $H_n$ is the Hermite polynomials, $N_n$ is the normalization constant.  $N_0=1/\sqrt{L l_{\rm B}\sqrt{\pi}}$, where $L$ is the length of the periodic potential along the $x$-direction, and $l_{\rm B}$ is the magnetic length. The center of the cyclotron orbit lies at $y_{k}=kl_{\rm B}^2$, where $k$ is the LL wavevector along the $x$-direction. For such a state the first order perturbation correction to the $n^{\rm th}$ LL is given by 
\begin{equation}
H^{\prime}_n(k,k^{\prime}) = \langle \psi_{n,k}|U({\bf r})| \psi_{n,k'}\rangle,
\end{equation}
where $U({\bf r})$ is the periodic potential.  Without loosing generality, we assume that the 1D potential spans along the $x$-direction, and thus it can be expanded in the Fourier basis as $U(x)=1/\sqrt{L} \sum_{{\bf q}}U(q)e^{iqx}$, where $q$ is determined by the nature of the periodic potential. We have a 1D square-like potential with a width $W_w\approx W_b=L$, where $W_b$ is the width of $U=0$ intermediate region. However, the potential edge is a smooth function in $x$ and and one can assume a sinusoidal form, governing discrete values of $q=2\pi n/L$. 

We can now compute the matrix element of the perturbed Hamiltonian as
\begin{eqnarray}
&&H{'}_n(k,k')
=\frac{N_n^2}{L}\sum_{q}\int_{-\infty}^{\infty}\int_{-\infty}^{\infty}{dxdy}~U(q) e^{i(k-k' + q)x}\nonumber\\
&&\qquad\qquad \times H^*_n\left(\frac{y-y_k}{l_B}\right)H_n\left(\frac{y-y_{k'}}{l_B}\right)~e^{-{\frac{(y-y_k)^2+(y-y_j)^2}{2l_{\rm B}^2}}}.
\label{perturbation1}
\end{eqnarray}
This equation is analogous to the tight-binding equation, and in this spirit, we refer the tight-binding hopping amplitude $t_n(k,k')=H{'}_n(k,k')$ between the localized cyclotron orbits located at $y_k=kl_B^2$, and $y_{k'}={k'}l_B^2$. In Eq.~\eqref{perturbation1}, the $x$-integral gives the momentum conservation condition $k'=k+q$. The $y$-integral gives different results for different LLs, and for the zeroth LL of present interest, we obtain
\begin{equation}
t_0=\frac{1}{L}\sum_q U(q) e^{-(y_k-y_{k+q})^2/4l_{\rm B}^2}=\frac{1}{L}\sum_q U(q) e^{-\frac{(ql_B)^2}{4}}.
\label{TB}
\end{equation}
The absolute value of $t_0$ depends on $U(q)$. It depends linearly on the potential strength, as also obtained experimentally (see Fig.~6(c)). $t_0$ however increases exponentially on magnetic field, i.e, $t_0\propto e^{-\lambda/B}$, where $\lambda=\hbar q^2/4e$. 

In the second quantization language, the effective Hamiltonian becomes
\begin{eqnarray}
H'_n=\sum_{i,j}t_n c_{i}^\dagger c_{j}+h.c.
\end{eqnarray}
where $i,j$ denote the degenerate LLs with cyclotron orbits located at $y_i$ and $y_j$. By doing the Fourier transformation $(c_j=1/\sqrt{L}\sum e^{i p r_j}c_p)$, we obtain a dispersion  of the LL along the $x$-direction as
\begin{eqnarray}
E'_n(p)=2t_n\cos(pL),
\end{eqnarray}
where $L$ is the periodicity of the superlattice potential. The total tight-binding energy dispersion of the LL is thus 
\begin{eqnarray}
E_n (p) = {\rm{sign}}(n)\sqrt{|n|}\hbar\omega_c  + 2t_n\cos(pL). 
\end{eqnarray}
The bandwidth of the above dispersion is $4|t_n|$. The DOS at a given energy $\epsilon$ is defined as $\rho_n(\epsilon)\propto \sum_{p}\frac{1}{v_p}\delta(\epsilon-E_n(p))$, where the group velocity $v_p=\partial E_n(p)/\hbar\partial p=-2t\sin{(pL)}$. $v_p$ vanishes at $p=0,\pm\pi/L$, i.e. at the bottom and top of the bands. Therefore, the original LL acquires sharp peaks at the top and bottom of the dispersion in a 1D periodic lattice. The energy gap between the DOS peaks is $E_g^n=4|t_n|$. As mentioned in Eq.~\eqref{TB}, $t_n$, and hence $E_g^n$ increases linearly with $U$, but exponentially with the magnetic field $B$, both of which are consistent with experiments, as discussed in Fig.~6. The splitting decreases as we move to higher LL.  

In Fig.~6, we show the DOS as a function of energy for graphene at a given magnetic field $B=6$T, for the periodic potential $U=0$, (Fig. 6a) and $U=50meV$ (Fig. 6b). Also including a (experimental) broadening of $\sim$ 30 meV, we notice that $n=0$ LL broadens for small superlattice potential, and then it splits into two for large superlattice potential for $U<\hbar\omega_c$. For superlattice potential comparable to the LL gap $\hbar\omega_c$, scattering between different LL becomes large, and higher LLs start mixing with each other, which has been experimentally observed as shown in S8. In Fig. 6c we compare the theoretical splitting (red line) of $N=0$ LL with the experimental one as a function of superlattice potential, where one can clearly see the qualitative agreement between the experimental results with the theoretical one. We believe the mismatch is due to the overestimated values of the barrier potential.

\section{Conclusion}
In summary, we have demonstrated for the first time, quantum capacitance measurements in 1D graphene superlattice. We have observed band structure modification as a function of 1D superlattice potential strength at zero magnetic field is consistent with the theoretical prediction. At higher superlattice potential we have observed a large splitting ($\sim$ 150 meV) of the zero Landau level. The renormalization of the LL spectrum due to 1D superlattice potential is compared with our theoretical calculation. 
\section{Acknowledgements}
A.D thanks DST nanomission (DSTO-1470 and DSTO-1597) for the financial support.
\bibliography{ref_grid}{}

\onecolumngrid
\newpage
\thispagestyle{empty}
\mbox{}


\section*{Supplementary material}

# Supplementary Information

Manabendra Kuiri[1], Gaurav Kumar Gupta[1], Yuval Ronen[2,3], Tanmoy Das[1], and Anindya Das[1]
[1]*Department of Physics, Indian Institute of Science, Bangalore 560012, India*
[2]*Braun Center for Submicron Research, Department of Condensed Matter Physics,
Weizmann Institute of Science, Rehovot 76100, Israel and*
[3]*Department of Physics, Harvard University, Cambridge, MA 02138, USA*

## S1. Device Fabrication

The device was fabricated using the following steps. At first a degenerately doped n++ Si/SiO$_2$ substrate with 300 nm oxide was taken and cleaned using Pirhana solution. The top SiO$_2$ (around 150 nm) was etched out using buffered HF solution, followed by rinsing in deionized water (DI) and blow dry using N$_2$ gas. A oxide thickness of $\sim 150$ nm was necessary to get fine well spaced line, which is usually difficult to achieve with 300 nm oxide. Using electron beam lithography grids were patterned followed by thermal deposition of Ti/Au (5nm/20nm). The scanning electron microscope image of such gold grids is shown in Fig. S1b. hBN-graphene-hBN stack was made using the pick up technique as described in [S1] and deposited on top of the prepatterned gold grids as shown in Fig. S1a. Here, graphene was chosen bigger than the top hBN, but smaller than the bottom hBN. This was necessary to make two probe contacts in graphene without etching it [S2] as shown in Fig. S1c. Source(drain) and topgate electrodes were patterned using another step of electron beam lithography followed by thermal deposition of Cr/Au (5nm/70nm). The optical image of final device with electrical contacts is shown in Fig. S1c.

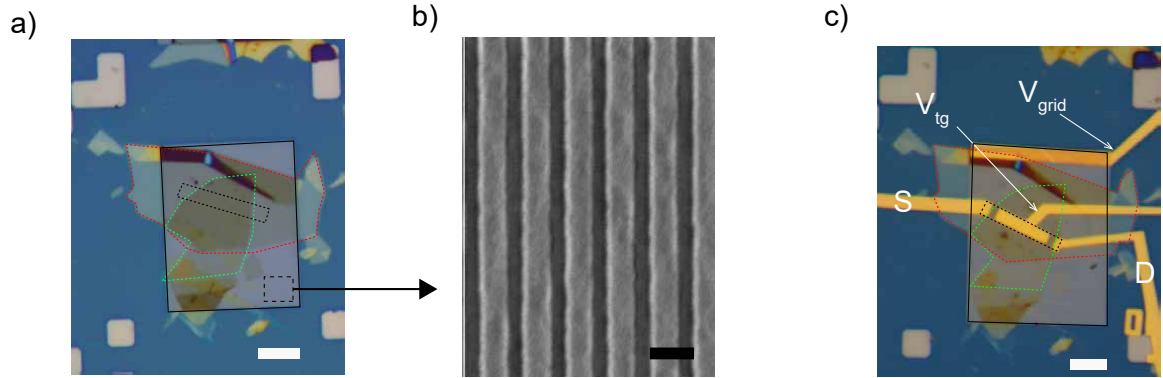

Fig. S1: (Color Online) (a) hBN-graphene-hBN stack placed on top of gold grids. Red dashed line shows the bottom hBN, black dashed line shows the graphene, green dashed line shows the top hBN and, black solid line shows the gold grids. Scale bar 4 $\mu m$. (b) Shows the SEM image of gold grids with $W_b = 60$ $nm$ and $W_w = 40$ $nm$, scale bar 100 nm.(c) Optical image of the device after making the electrical contacts. Scale bar 4 $\mu m$.

## S2. Estimation of hBN thickness

The thickness of top hBN was found using atomic force microscopy (AFM) imaging. The phase is shown in Fig.S2a, where we show the outline of top hBN, graphene and bottom hBN for clarity. The white dashed line marked in Fig. S2b shows the place where the height profile was taken. From the height profile shown in Fig. S2c thickness of top hBN was found $d_T \sim 10 - 11$ $nm$

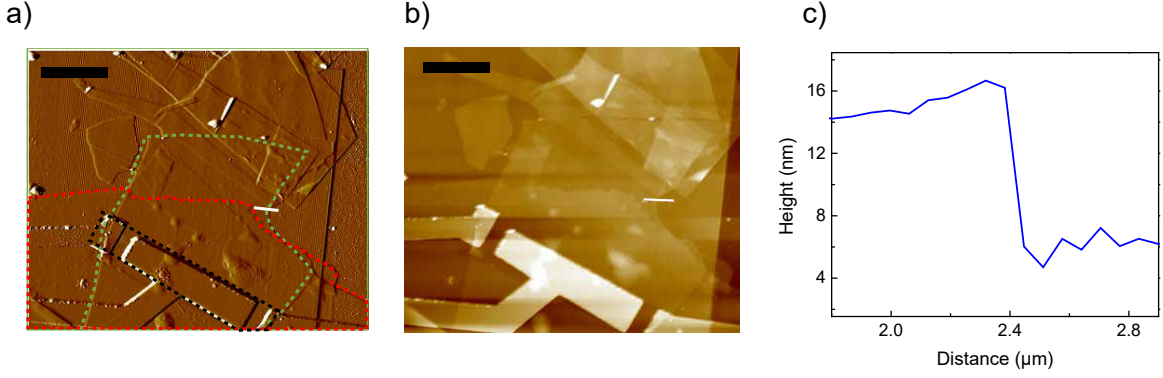

Fig. S2: (Color Online) AFM image of the measured device (a) and (b). (a) Shows the phase, where red dashed line shows the bottom hBN, black dashed line shows the graphene ,and green dashed line shows the top hBN. Scale bar 5 $\mu m$. (b) Shows the AFM image. The white dashed line shows the place where the height profile was taken. Scale bar 5 $\mu m$. (c) Height profile along the white line shown in (b)

### S3. Parasitic capacitance extraction

The total measured capacitance between the topgate and graphene is given by

$$C_t = \left(\frac{1}{C_g} + \frac{1}{C_q}\right)^{-1} + C_p \quad (1)$$

where $C_g \sim 55 fF$ is the geometric capacitance, $A = 18\mu m^2$ is the area of the device, $C_q = Ae^2 \frac{dn}{d\mu}$ is the quantum capacitance, $\frac{dn}{d\mu} = \frac{2|\mu|}{\pi(\hbar v_F)^2}$ is the DOS, $v_F$ is the Fermi velocity and $C_p$ is the parasitic capacitance. Now in order to compare with the experimental data, we convert Fermi energy ($\mu$) into gate voltage ($V_{tg}$). Noting that $eV_{tg} = ne^2/C_g + \mu$, and $\mu = \hbar v_F \sqrt{\pi n}$, we can rewrite

$$V_{tg} = \frac{\mu^2 e}{\hbar^2 v_F^2 \pi C_g} + \mu/e$$

. Using the values of $C_g$ and $C_q$, we calculate theoretical $C_t$ as a function of topgate voltage $V_{tg}$. The red curve in Fig. S3 shows the theoretical $C_t$ as a function of $V_{tg}$ for $C_g \sim 55\ fF$ with $v_F = 1.2 \times 10^6 m/s$. The experimental curve (blue) was matched with the theoretical curve and the parasitic capacitance was found to be $C_p \sim 4.8\ fF$.

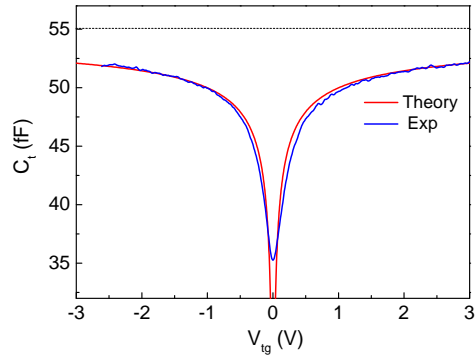

Fig. S3: (Color Online) (a) Total capacitance $C_t$ as a function of topgate voltage $V_{tg}$ for $V_{grid} = 0V$. The blue curve shows the experimental data with $C_p \sim 4.8\ fF$ subtracted , and the red curve shows the theoretical curve using Eq.(1),with $C_g = 55\ fF$, $v_f = 1.2 \times 10^6 m/s$, $A = 18\mu m^2$. The black dashed line shows $C_g \sim 55\ fF$.

### S4. Area calculation

The area was calculated by using the Scanning electron microscope (SEM) image. The graphene part was underlined as with white dashed line as shown in Fig. S4. Using Raith Eline software, the image was imported and the area was calculated (Black dashed line in Fig. S4). The topgate area was found to be $A = 18 \mu m^2$

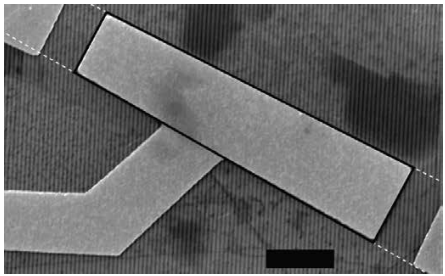

Fig. S4: (Color Online) (a) Shows the scanning electron microscope image of the device. The black dashed lines shows the part of graphene where the area was calculated. The topgate area was found to be $A = 18 \mu m^2$. The angle between graphene and the gold grids was found to be 60 degrees. The white dashed lines shows the edge of graphene. Scale bar 2 $\mu m$.

### S5. Barrier height calculation

The barrier height was estimated by considering the net doping in the regions. Region 1 ($W_w$ - barrier well between the gold pillars), where only the top gate control the density. The region 2 ($W_b$ - on top of the gold pillar), where we have combination of grid gate and topgate to control the density as shown schematically in Fig. S5b. The charge density induced by the topgate is given by $n_{tg} = \frac{C_{tg}V_{tg}}{e}$ and the induced charge density under the gridgate (gold pillar in Fig. S5b) region is given by the algebraic sum of densities due to both gates $n_{grid} = \frac{C_{grid}V_{grid}+C_{tg}V_{tg}}{e}$, where $C_{grid}(C_{tg})$ is the capacitance per unit area due to the grid gate (topgate), and $e$ is the electronic charge. $V_{grid}(V_{tg})$ are the grid(topgate) voltages measured relative to the Dirac point. In case of graphene $E_F = \hbar v_F \sqrt{\pi n}$, where $v_F$ is the Fermi velocity, $n$ is the number density of charge carriers, and $E_F$ is the Fermi energy measured relative to the Dirac point. The potential $U$ defined by both gates $V_{tg}$ and $V_{grid}$ is thus given by [S3]

$$U = \sqrt{\pi}\hbar v_F \left( \sqrt{\frac{C_{grid}\Delta V_{grid} + C_{tg}\Delta V_{tg}}{e}} - \sqrt{\frac{C_{tg}\Delta V_{tg}}{e}} \right) \quad (2)$$

$C_{tg}$ per unit area is given by $\frac{\epsilon_0 \kappa}{d_t}$, where $\epsilon_0$ is permittivity in free space, $\kappa \sim 3.9$ is the dielectric constant of hBN, and $d_t \sim 11\ nm$ is the thickness of top hBN. The diagonal line in Fig. 2a represents the charge neutrality point.

$$C_{grid}\Delta V_{grid} = C_{tg}\Delta V_{tg} \quad (3)$$

The slope of the diagonal line in Fig. 2a gives the ratio of topgate to grid gate capacitance per unit area.

$$\frac{C_{tg}}{C_{grid}} \sim 2.1 \quad (4)$$

$C_{grid}$ per unit area is given by $\frac{\epsilon_0 \kappa}{d_{grid}}$, where $d_{grid}$ is the thickness of the bottom hBN, from Eq.(4) knowing $d_t \sim 11\ nm$, $d_{grid}$ turns out to be $\sim 22\ nm$. Using the above values of $C_{tg}(C_{grid})$, the barrier height ($U$) was calculated using Eq.(2). Fig. S5c shows the colorplot of the barrier height ($U$) as a function of experimental range of $V_{grid}$ and $V_{tg}$ values.

The barrier height quoted in the main text for $V_{grid} = 1V$, $1.5V$, $2.0V$, $2.5V$ was calculated by taking the line cut of Fig. S5c for a fixed value of $V_{grid}$, and then taking the average $U$ between the topgate values where the capacitance modulation was observed. For instance in Fig. S6 for $V_{grid} = -2.5V$, the capacitance modulation occurs for $V_{tg} \sim (0.35 - 1.5)V$, $U$ was calculated by taking the average value of $U$ over this range of topgate voltage.

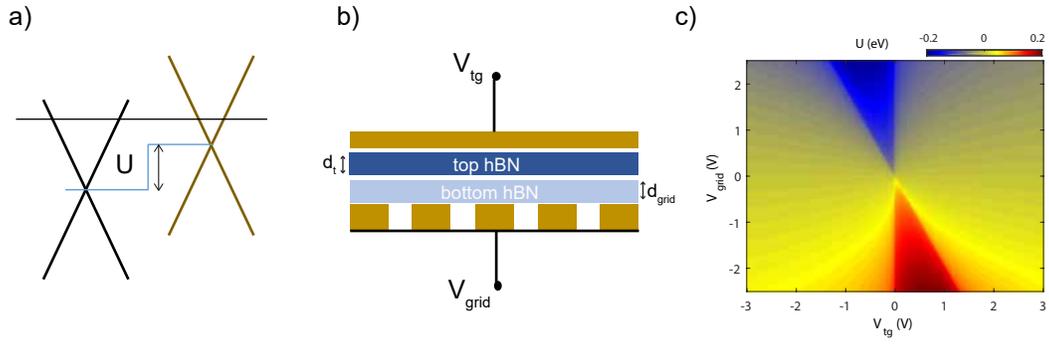

Fig. S5: (Color Online) (a) Potential barrier created by the two gates. (b) Schematic of the gates configuration where $d_{grid}$ is the thickness of the bottom hBN, $d_t$ is the thickness of the top hBN (b) Potential profile of the device as a function of $V_{tg}$ and $V_{grid}$

### S6. Capacitance 2D plot for B=0

The measured total capacitance $C_t$ as a function of topgate voltage $V_{tg}$ for different value of grid voltage is shown in Fig. S6.

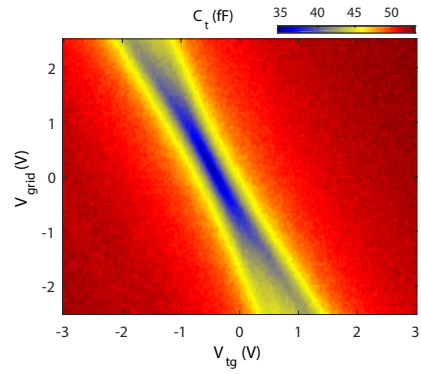

Fig. S6: (Color Online) (a) Shows the total capacitance measured ($C_t$) as a function of $V_{tg}$.

## S7. Magneto capacitance Data

The magneto capacitance data was recorded by keeping the $V_{grid}$ fixed, and sweeping the topgate voltage ($V_{tg}$) for different values of magnetic field (B) and measuring the total capacitance ($C_t$). Fig. S6 shows the colorplot of $C_t$ vs $V_{tg}$ for different magnetic field for $V_{grid} = -1V, -1.5V, -2.0V, -2.5V$.

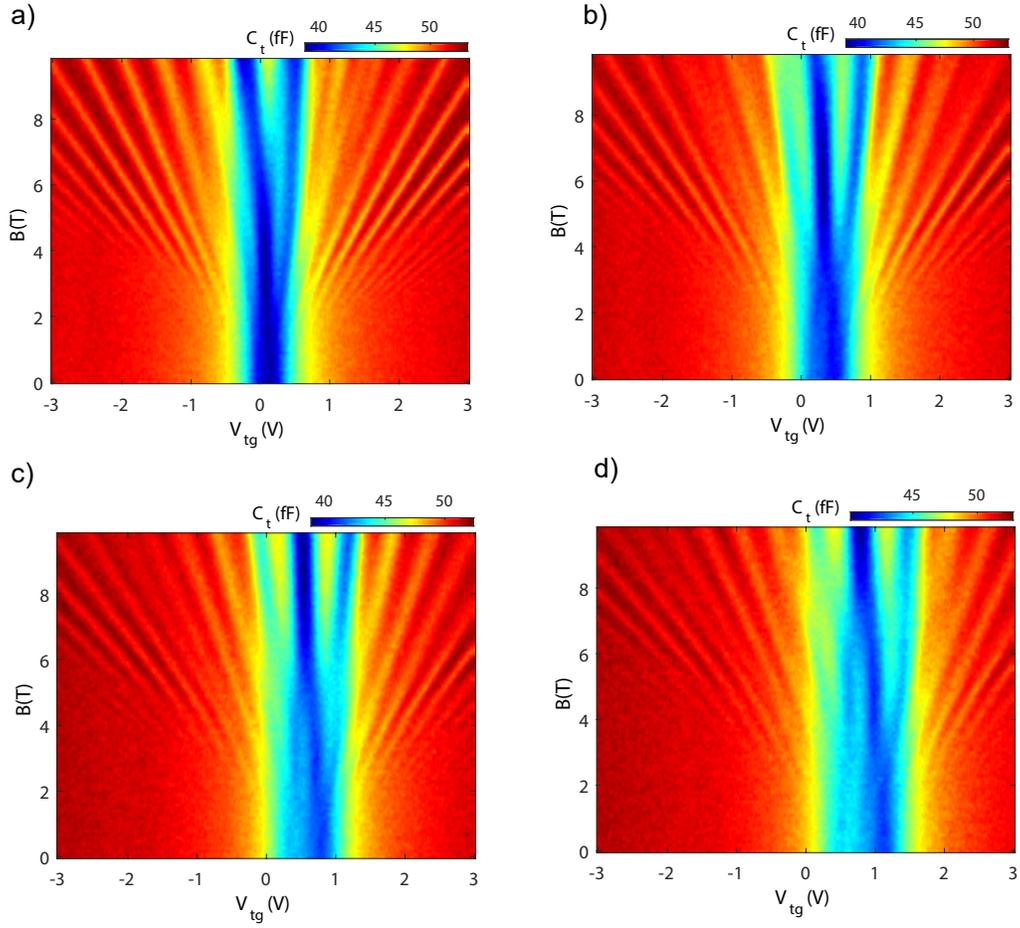

Fig. S7: (Color Online) Total capacitance measured ($C_t$) as a function of $V_{tg}$ for $V_{grid}$ = - 1V (a) -1.5V (b) -2.0V (c) and - 2.5V (d) for different values of magnetic field B

**S8. Magneto-capacitance Data for high superlattice potential** In Fig. S8 we have plotted the magneto-capacitance data for $V_{grid} = -2.5V$ with Fermi energy. In the main text we have already shown the data for other grid voltages. As shown in the main text (Fig. 5) with the application of 1D superlattice potential the zeroth L.L gets splitted. However, at higher $U \sim 165 meV$ (Fig. S8) the splitted zeroth L.L starts interacting with nearest L.Ls as the potential energy is comparable to the energy gaps between the L.Ls.

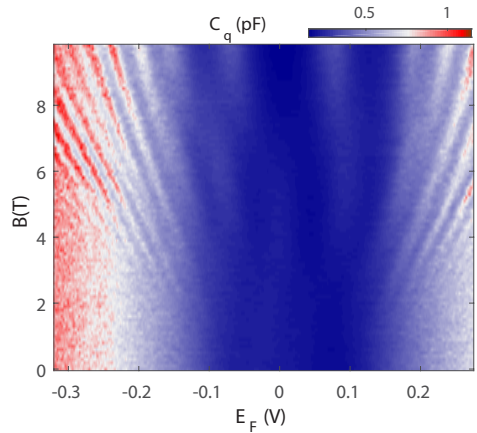

Fig. S8: (Color Online) $C_q$ vs $E_F$ as a function of magnetic field for $V_{grid} = -2.5V$


\end{document}